# Improved Charge Carrier Dynamics of $CH_3NH_3PbI_3$ Perovskite Films Synthesized by Means of Laser-Assisted Crystallization

*Ioannis Konidakis*[1], Temur Maksudov[2,3], Efthymis Serpetzoglou[1,4], George Kakavelakis[2,3], Emmanuel Kymakis[2], Emmanuel Stratakis*[1,3]*

[1]Institute of Electronic Structure and Laser (IESL), Foundation for Research and Technology-Hellas (FORTH), 71110 Heraklion, Crete, Greece.

[2]Center of Materials Technology and Photonics, Electrical Engineering Department, Technological Educational Institute (TEI) of Crete, 71004 Heraklion, Crete, Greece.

[3]Department of Materials Science and Technology and [4]Physics Department, University of Crete, 71003 Heraklion, Crete, Greece.

**Corresponding Authors**

*+30 2810-391835. E-mail: ikonid@iesl.forth.gr

*+30 2810-391274. E-mail: stratak@iesl.forth.gr

**Abstract**

Although it has been recently demonstrated that the laser-assisted (LA) crystallization process leads to the formation of perovskite absorber films of superior photovoltaic performance compared to conventional thermal annealing (TA), the physical origin behind this important discovery is missing. In this study, $CH_3NH_3PbI_3$ perovskite thin films have been synthesized by means of LA and TA crystallization on the surface of two hole transport layers (HTL) namely poly(3,4-ethylenedioxythiophene)-poly(styrenesulfonate) (PEDOT:PSS) and poly(triarylamine) (PTAA). A systematic study of the effect of laser irradiation conditions on the crystalline quality and morphology of the perovskite films was performed via scanning electron microscopy, X-Ray diffraction and absorption spectroscopy. Meanwhile, time-resolved transient absorption spectroscopy under inert atmosphere conditions was used to evaluate the carrier transport dynamics. It is found that for the PEDOT:PSS/$CH_3NH_3PbI_3$, structures the LA process resulted to perovskite layers of larger grains, faster charge carrier extraction properties and slower bimolecular recombination, when compared to TA. On the contrary, the LA-assisted formation of the PTAA/$CH_3NH_3PbI_3$ heterostructures leads to extensive presence of residual $PbI_2$ and thus inferior performance and charge carrier dynamics.

## 1. Introduction

During the last decade organic and inorganic lead halide perovskites have been the main subject of numerous scientific studies due to their enormous potential for being the main component of next generation photovoltaics.[1-6] At the same time, their impact in other important photonic devices beyond photovoltaics has been also well-established, predominantly in the fields of light-emitting devices,[6-9] photodetectors,[10-12] lasers,[13-15] and various sensing platforms.[16-18] In terms of photovoltaic technology the progress over the recent years is emphatic, as from the introduction of perovskite solar cells (PSCs) their power conversion efficiency (PCE) in typical small area devices of ~1 $cm^2$ has been enhanced from ~4% up to ~22%,[1-3,19] This remarkable progress makes perovskites outstanding candidates for being the dominant absorber material in next-generation commercially available photovoltaic devices. However, for the realization of this major milestone, more scientific knowledge is still necessary in order to maximize further the PCE while resolving important stability issues upon humidity[20,21] and prolonged light exposure.[22,23]

In order to tackle the aforementioned challenges two main scientific strategies have been followed over the years. Based on the first, particular attention is given in the selection of suitable conductive polymer materials for the hole transport layer (HTL) and electron transport layer (ETL) of the PSC.[3-5,24,25] Such selection is important as it is proven that the interactions between the perovskite film and the transport layers are critical for the overall device performance. Namely, particular emphasis is given on diminishing ion migration within perovskite lattice, as well as, reducing light-activated charge states.[23,26,27] Doping the transport layers polymers with appropriate ingredients that increase conductivity and extend operating lifetime has also been successfully attempted.[28-30] Rather differently, the second approach focuses on the composition and synthesis of the perovskite itself as it is generally accepted that the crystalline quality of the hybrid absorber film plays a vital role in the PSC performance.[2-5,31-35] Within this frame, several methyl ammonium lead single- and mixed-halide perovskites of the families $CH_3NH_3PbX_3$ and $CH_3NH_3PbI_{3-x}X_x$, where X can be Cl or Br, have been used for the fabrication of advanced PSCs.[36,37] At the same time the influence of both one-step and two-step depositions has been explored,[1,38,39] as well as, the formation of crystalline absorber films by means of solution and vapor deposition processes.[7,13,40-42]

Along these recent progressive directions, the importance of employing laser-assisted (LA) crystallization for the development of advanced perovskite films is highlighted.[8,34,35,43] Namely, it is reported that when LA crystallization is followed instead of the typical thermal annealing (TA), the PCE of the perovskite photovoltaic is enhanced.[34,35] While the exact physical origin of this enhanced electrical performance in terms of charge transfer processes has not been identified yet, it was clearly shown that the rapid LA crystallization results to the formation of perovskite films with larger grains, improved density and better homogeneity.[34,35] It is worth to note also that the employment of laser beams poses numerous additional advantages in terms of synthesizing micro-patterned perovskite structures through direct laser writing (DLW) that expand functionality further towards the realization of prototype perovskite devices beyond photovoltaic applications.[44-47]

In this study, we synthesize two sets of $CH_3NH_3PbI_3$ perovskite films via LA and TA crystallization, on two different types of HTL polymers. The first set is synthesized on hydrophilic surfaces of poly(3,4-ethylenedioxythiophene)-poly(styrenesulfonate) polymer (PEDOT:PSS), whereas the second series of films is crystallized on more hydrophobic surfaces of poly(triarylamine) polymer (PTAA). In order to shed light on the charge extraction processes and related transient decay dynamics, all fabricated HTL/$CH_3NH_3PbI_3$ architectures are examined by means of femtosecond (fs) time-resolved transient absorption spectroscopy (TAS). TAS is a very powerful technique for providing important insights on the charge carrier dynamics such us electron-hole injection times and exciton recombination rates that are well-known to be directly correlated with the performance of perovskite[25,48-57] and organic[58-60] photovoltaic devices.

All TAS measurements were performed entirely under inert conditions, immediately after the perovskite film synthesis, while it is the first time to the best of our knowledge that such studies are made on perovskite films prepared via LA crystallization. It is found that the LA process gives rise to faster charge carrier dynamics in the case of PEDOT:PSS/$CH_3NH_3PbI_3$ system. These findings account plausibly for the recently reported enhanced PCEs of the photovoltaic absorbers prepared by means of LA crystallization.[34,35] On the contrary the LA process is found to be considerably less effective for the PTAA/$CH_3NH_3PbI_3$ system. Our findings shed light on the influence of the perovskite crystallization process combined with the HTL layer used on the structural morphology and carrier transport properties of perovskite photovoltaic layers.

## 2. Experimental

### 2.1 Sample fabrication

All HTL/$CH_3NH_3PbI_3$ configurations of this study were prepared on 2 x 1.5 $cm^2$ glass substrates (Naranjo Substrates) that are coated with pre-patterned indium tin oxide (ITO). Before the HTL deposition, a three-step rinsing procedure with de-ionized water, acetone, and isopropanol was followed in order to ensure that all ITO/glass surfaces were completely free of impurities.[25,29] Following this cleaning process, the substrates were positioned within an ultraviolet ozone cleaning apparatus in order to improve ITO-coated glass surface hydrophilicity while ensuring the complete removal of organic residues. Following this treatment, the ITO/glass surfaces were ready for the deposition of the HTL polymers, i.e. PEDOT:PSS or PTAA. For samples where the former polymer is employed as HTL, a ~30 nm thick layer was spin-casted on the ITO surface from an aqueous solution (4000 rpm, 60 s). Afterwards, the sample was baked at 120 $^oC$ for 15 min under lab ambient conditions, followed by 30 more minutes of heating at the same temperature within a nitrogen-filled glove box.[25] In the case of PTAA HTL, a solution of PTAA powder in toluene (7 mg $mL^{-1}$) doped with 1.5% wt tetrafluoro-tetracyanoquinodimethane was prepared. For the deposition of PTAA HTL the latter solution was spin-coated on the ITO surface at a rate of 4000 rpm for 35 s. After completing HTL deposition, all substrates were allowed to cool down to room temperature (RT).

The synthesis of the $CH_3NH_3PbI_3$ perovskite films was performed by means of either conventional thermal annealing or by laser-assisted (LA) crystallization. For both fabrication protocols the perovskite precursors were introduced on the HTL substrate by executing a standard solution-based two-step procedure that makes use of two separate anhydrous solutions within a nitrogen-filled glass box.[25] The first solution is prepared by dissolving lead iodide ($PbI_2$) in dimethylformamide (450 mg $mL^{-1}$) and stirring overnight at 75 $^oC$, whereas the second solution is prepared by adding methylammonium iodide ($CH_3NH_3I$) in isopropanol (45 mg $mL^{-1}$) and stirring at RT. In the first step of the procedure, the $PbI_2$ solution is spin-coated on the HTL and heated at 100 $^oC$ for 10 min. Up to this point the procedure followed for the perovskite film synthesis is identical regardless the selected annealing protocol, i.e. thermal or LA. For the thermal annealing synthesis route, in the second step the $CH_3NH_3I$ solution was casted on the surface of $PbI_2$ and heated at 100 $^oC$ for another 30 min. This completed the preparation of HTL/$CH_3NH_3PbI_3$ samples synthesized by means of typical thermal annealing crystallization.

On a quite different manner, in the second step of the LA crystallization method the $CH_3NH_3I$ solution was casted without any heating. Rather differently, upon spin-coating the crystallization of $CH_3NH_3PbI_3$ perovskite films was instantly achieved by irradiating the precursor solutions with a continuous wave laser operating at 532 nm. Figure 1a shows a schematic representation of the LA crystallization experimental setup. Namely, a 10x microscope objective was used to focus the laser beam through a spatial filter, whereas after the filter an optical lens collected the so-formed uniform beam and guided it perpendicularly on the sample surface. The whole process took place under inert conditions, with the sample remained enclosed into a specially designed chamber evacuated and purged with nitrogen gas (Figure 1b). This configuration allowed the realization of homogeneous $CH_3NH_3PbI_3$ perovskite film spots of ~5 mm in diameter. The laser power density ranged from 6.4 to 11.4 W $cm^{-2}$, whereas the laser exposure time was kept constant to 20 s for all LA crystallizations. All fabricated HTL/$CH_3NH_3PbI_3$ sample configurations and the corresponding synthesis parameters are summarized in Table 1.

### 2.2 Characterization techniques

Following sample fabrication all HTL/$CH_3NH_3PbI_3$ configurations were characterized by TAS, performed also under inert conditions with the sample remained sealed inside the chamber throughout the measurements. TAS scans were conducted on a Newport (TAS-1) transient absorption spectrometer while using a 1026 nm pulsed laser source beam generated from an Yb:KGW-based laser system (PHAROS, Light Conversion), i.e. with 170 fs pulse duration and 1 kHz repetition rate. The configuration of TAS is shown schematically in Figure S1, while the experimental details of the measurements are described in the supporting information (SI) and elsewhere.[25] For the present study, all TAS measurements were performed with a pump beam energy of 1 mJ $cm^{-2}$, i.e. enough to cause the desired excitations without damaging the perovskite films.[25]

A field emission scanning electron microscope (JEOL, JSM-7000F) was employed for the thorough morphological characterization of the perovskite films, whereas their crystalline quality was studied by means of X-ray diffraction (XRD). For the XRD measurements an X-Ray Rigaku (D/max-2000) diffractometer was used while being operated with a continuous scan of Cu Ka1 radiation with λ equal to 1.54056 Å. All XRD scans were captured with a scan rate of $0.1^{o}$ $s^{-1}$. Finally, a PerkinElmer UV/VIS (Lambda 950) spectrometer was used to investigate the optical absorption properties of the fabricated HTL/$CH_3NH_3PbI_3$ films over the wavelength range of 430-820 nm.

## 3. Results and discussion

### 3.1 PEDOT:PSS/$CH_3NH_3PbI_3$ film architectures

Figure 2a shows a scanning electron microscope (SEM) image of the thermally annealed $CH_3NH_3PbI_3$ film on PEDOT:PSS, while Figures 2b-f present the corresponding photos of $CH_3NH_3PbI_3$ films synthesized on the same polymer by means of LA crystallization. Inspection of Figures 2b-f reveals that the morphology of the so-formed PEDOT:PSS/$CH_3NH_3PbI_3$ films strongly depends on the employed laser power density. Figure 3 depicts a plot of the maximum perovskite grain size as a function of laser power density, whereas the maximum grain size of the thermally annealed sample is also marked for the sake of comparison. The maximum grain size of each crystalline film was retrieved after thorough analysis of SEM profiles with the aid of sophisticated computer software.[61] Two striking observations are noted. Firstly, the initial increase of laser power density up to 8.9 W $cm^{-2}$ assists the formation of larger perovskite grains, whereas further rise of laser power appears to be harmful for the grains size of the crystalline films. Secondly, it is revealed that upon applying optimum laser power density, i.e. 8.9 W $cm^{-2}$, the LA crystallization procedure results to the formation of crystalline films with considerably larger grains when compared to the ones synthesized by conventional thermal annealing. Such findings are in agreement with recent studies that highlight the achievement of enhanced quality perovskite absorber films upon using LA crystallization methods.[34,35] Notably however, for the highest laser power density of 11.4 W $cm^{-2}$ severe decomposition of the $CH_3NH_3PbI_3$ perovskite is observed, i.e. resulting to inhomogeneous films (Figure 2f). This renders sample LA-PD-5 inappropriate for further characterization.

Aiming to explore further the crystalline nature of the obtained PEDOT:PSS/$CH_3NH_3PbI_3$ films XRD investigation was performed and the corresponding normalized patterns are shown in Figure 4a, while the raw data spectra are included in the SI as Figure S2. All XRD profiles of Figure 4a exhibit two characteristic peaks at 14° and 28.4°, corresponding to the (110) and (220) lattice planes of the $CH_3NH_3PbI_3$ perovskite, respectively.[29,34,35,41,50] An additional XRD feature at 12.6° is also noted and attributed to the presence of residual or unreacted $PbI_2$. Notably, the $PbI_2$ XRD peak is present for all perovskite films independently of the followed synthesis protocol, i.e. thermal annealing or LA crystallization, while such $PbI_2$ excess has been reported in several occasions to be harmless for the composition and optical properties of the absorber films and beneficial for their electrical performance.[32,33,39,62,63] For shedding light on the optical properties of PEDOT:PSS/$CH_3NH_3PbI_3$ films, the absorbance spectra of all configurations are measured and shown in Figure 4b. In agreement to previous reports, all absorbance spectra of Figure 4b exhibit two distinct features at ca. 490 nm and ca. 760 nm that emerge from the transitions of the dual valence band structure of $CH_3NH_3PbI_3$ films.[48-51,57] Inspection of Figure 4 reveals also that all films absorb readily in the visible region below 750 nm, however, with the exception of sample LA-PD-4, the LA crystallization perovskite films exhibit more pronounced absorbance when compared to the reference thermally annealed sample TA-PD. In a recent study by Wehrenfennig et al.[51] it was recognized that better absorbance in the visible range, and thus reduced amount of scattering, may be evidence for the formation of absorber films with enhanced quality and improved electrical performance.

In order to investigate the photovoltaic potential of PEDOT:PSS/$CH_3NH_3PbI_3$ films of this study, while revealing the underlying mechanisms of the charge carrier transport processes and corresponding dynamics, the so-formed film architectures were examined by TAS. Figure 5a displays typical ΔOD versus wavelength plots of the thermally annealed PEDOT:PSS/$CH_3NH_3PbI_3$ film (TA-PD) at various time delays following photoexcitation at 1026 nm with a pump fluence of 1 mJ $cm^{-2}$. Figure 5b presents the corresponding TAS spectra of a LA crystallization sample (LA-PD-2), i.e. typical spectra for all LA-PD samples. In both figures the dominant ΔOD peak at ca. 750 nm is attributed to the transient photoinduced bleaching of the perovskite absorber film, while a less pronounced photoinduced transient absorption feature is also noted and found to diminish with time in the range of 590-700 nm. For revealing maximum information on the relaxation dynamics and charge carrier transport within the so-formed PEDOT:PSS/$CH_3NH_3PbI_3$ architectures the transient decay kinetics of the photoinduced bleaching were thoroughly examined by adopting both exponential and polynomial fitting analyses. As explained previously,[25] the exponential fitting model allows us to determine the crucial time components of the charge carrier transport processes between the employed HTL polymer and the perovskite film,[25,50,64] whereas the higher order polynomial fitting provides insights on the kinetic rates of vital recombination processes that occur within the absorber films.[49,51,53-55,65,66]

Figure 5c shows the three-exponential fitting for the reference TA-PD sample, while Figure 5d presents the corresponding fittings for the four LA crystallization samples. In all cases the three-exponential fittings are based on the equation $y = y_o + A_1 \exp(-x/\tau_1) + A_2 \exp(-x/\tau_2) + A_3 \exp(-x/\tau_3)$, whereas the adjusted R squared ($R^2$) values exceeded 0.996, i.e. indicative of good quality fittings. Table 2 summarizes the obtained kinetic fit parameters for all PEDOT:PSS/$CH_3NH_3PbI_3$ architectures, i.e. $\tau_1$, $\tau_2$, and $\tau_3$. In particular, the first time component, $\tau_1$, is related to the charge carrier trapping between the perovskite grain boundaries and perovskite/PEDOT:PSS interfaces.[25,50] Inspection of Table 2 shows that $\tau_1$ decay component is found to be considerably faster for all LA crystallization samples when compared to the thermally annealed TA-PD film. Indicatively, all LA crystallization configurations exhibit at least two times smaller $\tau_1$ components, whereas for sample LA-PD-3 an up to six times faster $\tau_1$ is obtained, i.e. 62 ps compared to 10 ps, respectively. The physical significance of these findings is that in the case of LA crystallization absorber films shorter times are required for an excited electron to move from the conduction band to the trap states, and thus, quicker traps filling occurs. Consequently, larger splitting in the quasi-Fermi energy levels is achieved, which in combination with the more efficient free charge carrier injection to the HTL polymer, favor significantly the electrical characteristics of the absorber films.[25,67,68]

Furthermore, the second time component $\tau_2$ is assigned to the electron-hole injections from the perovskite film into the HTL polymer.[25,50,64] As is the case for $\tau_1$, data of Table 2 reveals also considerably quicker $\tau_2$ times for the LA crystallization films in comparison to the reference TA-PD architecture prepared by following the typical thermal annealing route. Namely, $\tau_2$ for sample TA-PD is 440 ps, whereas the corresponding $\tau_2$ times for LA-PD samples lie within the range of 111 to 224 ps. Strikingly enough, LA-PD-3 film exhibits a $\tau_2$ of 111 ps, which is four times faster when compared to the $\tau_2$ of the TA-PD sample. These findings strongly imply that the employment of LA crystallization techniques upon synthesis results to considerably faster hole extraction from the perovskite film into the PEDOT:PSS polymer. In one of our previous studies,[25] it was clearly demonstrated that the fast hole injection dynamics within the HTL/$CH_3NH_3PbI_3$ architecture correlate directly with the enhanced electrical performance of the PSCs of such configuration. Thus, the as-determined faster hole injection dynamics of the LA-PD samples, provide plausible reasoning for the advanced photovoltaic characteristics reported recently for PSCs that are fabricated by means of LA crystallization procedures.[34,35]

Finally, Table 2 includes the third long-life time component $\tau_3$, which is representative of the exciton recombination time. [25,50,64] In general, the primary recombination processes take place at two different time scales upon photoexcitation of the PEDOT:PSS/$CH_3NH_3PbI_3$ architectures. Namely, a fast process with lifetimes ranging from picoseconds to nanoseconds describes the electron-hole recombination kinetics of the perovskite films,[69] whereas, considerably slower trap-assisted recombination mechanisms on the nanosecond timescale occur at the grain boundaries and film interfaces.[69,70] While the latter processes lie out of the delay range employed for the TAS experiments of the present study, the exciton (electron-hole) recombination kinetics of the absorber films is readily depicted by $\tau_3$ (Table 2). Examination of Table 2 reveals that LA crystallization samples exhibit at least an order of magnitude slower $\tau_3$ time components, indicative of the presence of excitons with significantly longer lifetimes that are postulated to be beneficial in terms of their photovoltaic performance.[34,35]

Apart from the exponential fitting approach, we perform also the well-established polynomial fitting model based on the rate equation $dn(t)/dt = -k_3n^3 - k_2n^2 - k_1n$, where n is the charge carrier density and $k_1$, $k_2$, and $k_3$ are the rate constants corresponding, respectively, to trap-assisted recombination, bimolecular recombination, and Auger trimolecular recombination processes. [49,51,53-55,65,66] Figure S3 presents the polynomial fittings of the transient band edge bleach kinetics of PEDOT:PSS/$CH_3NH_3PbI_3$ films, whereas Table 3 lists the corresponding charge carrier rate constants. Among the three rate constants, the importance of $k_2$ in terms of the electrical performance of the perovskite absorber films was highlighted recently by Wehrenfennig et al.[51] Namely it was proposed that slower $k_2$ rates indicate longer free-charge carrier diffusion lengths within the films, which favor the efficiency of planar heterojuction PSCs based on such configurations. Inspection of Table 3 reveals that LA crystallization PEDOT:PSS/$CH_3NH_3PbI_3$ films exhibit an order of magnitude slower $k_2$ rates when compared to that of TA-PD sample, whereas sample LA-PD-3 exhibits the lower $k_2$ of all. We recall at this point that the latter sample exhibited also the fastest hole injection time among all architectures in question (Table 2), as well as, the largest grain size (Figure 3). Such findings provide a complete explanation for the recently reported enhanced photovoltaic performance of LA crystallization $CH_3NH_3PbI_3$ films.[34,35]

In summary, the findings of the present study reveal the combination of elements that render the LA-crystallized PEDOT:PSS/$CH_3NH_3PbI_3$ absorbers more advanced in terms of their photovoltaic potential. It becomes apparent that the employment of more rapid crystallization protocol produces perovskite grains of larger size, while accelerating the charge carrier injection to the HTL and elongating the lifetime of excitons. Moreover, the LA fabrication protocol assists the production of perovskite films with longer free-charge carrier diffusion lengths. Remarkably though, the benefits of LA crystallization process for synthesizing $CH_3NH_3PbI_3$ absorber films are apparent even for films that exhibit smaller grains when compared to the thermally annealed film, i.e. the LA-PD-1 and LA-PD-2 versus TA-PD. The former LA-processed samples exhibited better time decay components and recombination rate dynamics, despite the fact that their grain sizes were smaller than those of TA-PD sample (Figure 3). These findings emphasize further the superiority of the LA crystallization protocol in terms of achieving improved charge carriers collection efficiency between the absorber film and the HTL polymer,[71] i.e. PEDOT:PSS in this case.

### 3.2 PTAA/$CH_3NH_3PbI_3$ film architectures

As a second part of this study, we attempted to exploit the benefits of LA crystallization for the formation of $CH_3NH_3PbI_3$ films on the surface of PTAA polymer. This was prompted by our recent findings, where the employment of PTAA instead of PEDOT:PSS resulted to the formation of absorber films with better morphological characteristics and improved electrical performance.[25] Figure 6 shows SEM images of PTAA/$CH_3NH_3PbI_3$ films grown by means of conventional thermal annealing (Figure 6a) and by following LA crystallization synthesis at various laser power densities (Figures 6b-f). It becomes immediately apparent from Figure 6 that the thermally annealed sample (TA-PT) exhibits larger grains in comparison to all LA samples. Computerized analysis depicted in Figure 7 confirms this observation.[61] In particular, the maximum grain size is achieved at a laser power density of 8.9 W $cm^{-2}$, whereas further power augmentation reduces grain size until eventually causing film ablation. Thus, on the contrary to what was found for PEDOT:PSS/$CH_3NH_3PbI_3$ architectures, LA synthesis route appears to be disadvantageous in terms of forming perovskite films with larger grain sizes on the surface of PTTA polymer. In addition, Figure 8a presents the normalized XRD patterns of PTAA/$CH_3NH_3PbI_3$ films, while the corresponding raw data spectra are shown in the SI as Figure S2. All XRD profiles show the characteristics $CH_3NH_3PbI_3$ peaks at 14$^o$ and 28.4$^o$. However, a striking observation is that the XRD spectra of LA samples reveal the presence of extensive excess of $PbI_2$, particularly, for samples LA-PT-2, LA-PT-3, and LA-PT-4 that exhibited the larger grain sizes among the LA crystallization films. The absorbance spectra of PTAA/$CH_3NH_3PbI_3$ films (Figure 8b) appear similar without any obvious abnormalities.

TAS provides multiple insights on how the revealed peculiar morphological and composition characteristics of the LA crystallization PTAA/$CH_3NH_3PbI_3$ films affect their charge carrier extraction efficiency and the corresponding recombination rates. Figure 9a shows typical ΔOD versus wavelength plots of the thermally annealed PTAA/$CH_3NH_3PbI_3$ architecture (TA-PT), whereas Figure 9b presents typical TAS spectra of a LA crystallization sample, i.e. LA-PT-2. All plots exhibit the characteristic photoinduced bleaching peak at ca. 750 nm, which is found to attenuate with time as expected. Figure 9c depicts the transient band edge bleach kinetics and their three-exponential decay fitting for TA-PT sample, while Figure 9d illustrates the corresponding plots for LA crystallization samples LA-PT-1 and LA-PT-2. All time components from the exponential fitting analysis are listed in Table 2. In agreement to our previous findings,[25] for the thermally annealed architectures faster time components are revealed when the PTAA polymer is employed instead of PEDOT:PSS. Namely, $\tau_1$ reduces from 62 to 31 ps, whereas the critical electron-hole injection component $\tau_2$ drops from 440 to 375 ps. Moving to the LA crystallization PTAA/$CH_3NH_3PbI_3$ architectures, $\tau_1$ times for samples LA-PT-1 and LA-PT-2 are found equal to 64 and 57 ps, respectively, while the corresponding $\tau_2$ times are determined equal to 2256 and 1813 ps, respectively. Oddly enough, both samples LA-PT-1 and LA-PT-2 exhibit remarkably slower $\tau_1$ and $\tau_2$ components when compared to the thermally annealed sample. On a similar comparison, Table 3 reveals faster $k_2$ rates for both LA samples, as determined from the polynomial fitting analysis of the bleach kinetics data (Figure S3d). These findings imply that the LA crystallization PTAA/$CH_3NH_3PbI_3$ configurations exhibit less efficient charge extraction, while having shorter free-charge carrier diffusion lengths. The combination of these factors renders the LA assisted crystallization protocol unfavorable when compared to typical TA synthesis route.

Although it was possible to perform good quality three-exponential and polynomial fittings for samples LA-PT-1 and LA-PT-2, that was not the case for the remaining two PTAA/$CH_3NH_3PbI_3$ samples, i.e. LA-PT-3 and LA-PT-4. Attempting to perform typical fitting models to data of samples LA-PT-3 and LA-PT-4 resulted to poor fitting quality. Indicatively, the inset of Figure 9d shows the bleach kinetics of sample LA-PT-3 without any fitting. Instead a steep change in slope is noted in the vicinity of 420 ps. We postulate that the reason for this peculiar behavior lies in the extensive presence of $PbI_2$ phase in samples LA-PT-3 and LA-PT-4 (Figure 8a), that eventually leads to passivation of the perovskite grain boundaries, as well as, slower charge carrier trapping at the PTAA/$CH_3NH_3PbI_3$ interfaces.[50] In overall, TAS data confirm what was evidenced by SEM and XRD findings, namely, that LA crystallization of $CH_3NH_3PbI_3$ films on PTAA surface is less successful than the conventional thermal annealing treatment, i.e. on the contrary to what is the case for PEDOT:PSS where the employment of LA crystallization enhances the film quality and properties.

Finally, we consider the reasons that render the rapid LA crystallization protocol detrimental in terms of synthesizing $CH_3NH_3PbI_3$ films on the surface of PTAA polymer. It is known from atomic force microscopy (AFM) studies that the surface of PTAA polymer is considerably smoother when compared to PEDOT:PSS, i.e. average roughness values are found equal to 0.51 and 0.94 nm, respectively (SI and Figure S4). The surface of the former polymer is also slightly more hydrophobic than the PEDOT:PSS surface.[25] Both these factors contribute towards considerably less surface tension dragging force in the case of PTAA polymer.[24] While this might be beneficial for the perovskite film quality upon adopting the prolonged thermal annealing protocol with duration of 30 min, it appears than when the 20s-rapid and spontaneous photothermal heating of LA crystallization is followed, it causes partial vaporization of the volatile $CH_3NH_3I$/isopropanol solution, and thus, leaving behind over-stoichiometric amount of $PbI_2$, i.e. particularly for the higher employed laser power densities. This argument is further supported by the fact that photothermal heating upon LA crystallization is recognized to be critical for the quality and morphology of the so-formed perovskite films,[35,72] while also it strongly depends on the nature of HTL polymer substrate.[35] Rather differently, in the case of PEDOT:PSS/$CH_3NH_3PbI_3$ films, the pronounced surface tension dragging force effect of PEDOT:PSS prevents the excessive vaporization of $CH_3NH_3I$/isopropanol solution, and consequently, the amount of unreacted $PbI_2$ is limited and similar to that present in corresponding perovskite films prepared by following typical thermal annealing treatments.

## 4. Conclusions

$CH_3NH_3PbI_3$ perovskite films were synthesized by means of the laser-assisted crystallization and conventional thermal annealing processes on the surface of two HTL polymers, i.e. PEDOT:PSS and PTAA. For the case of the former polymer, it was found that the employment of the rapid LA crystallization technique resulted to perovskite grains of larger grains when compared to typical thermal annealing. More importantly, TAS studies performed under ambient conditions revealed, that upon following the LA crystallization route for the development of PEDOT:PSS/$CH_3NH_3PbI_3$ architectures, significantly faster charge carrier extraction properties and slower bimolecular recombination rates were attained. These findings render the LA fabrication procedure beneficial in terms of cost-effectively synthesizing perovskite absorber films of superior crystalline quality and electric performance. On the contrary, when the smoother and more hydrophobic PTAA polymer is used, the thermal annealing route was proved to be more successful for fabricating PTAA/$CH_3NH_3PbI_3$ structures with enhanced carrier transport dynamics, since the LA crystallization resulted to excessive presence of residual $PbI_2$.

**Acknowledgement**

The authors are grateful to A. Manousaki (IESL, FORTH) for her technical assistance with SEM studies, and K. Savva (IESL, FORTH) for supervising XRD experiments. This work was supported by the European Research Infrastructure NFFA-Europe, funded by EU's H2020 Framework Program for research and innovation under grant agreement number No. 654360.

**Supporting Information**

TAS setup and experimental details, XRD patterns (raw data), TAS polynomial fits, AFM experimental details, PTAA polymer surface characterization.

## Notes

The authors declare no competing financial interest.

## Figures and Tables

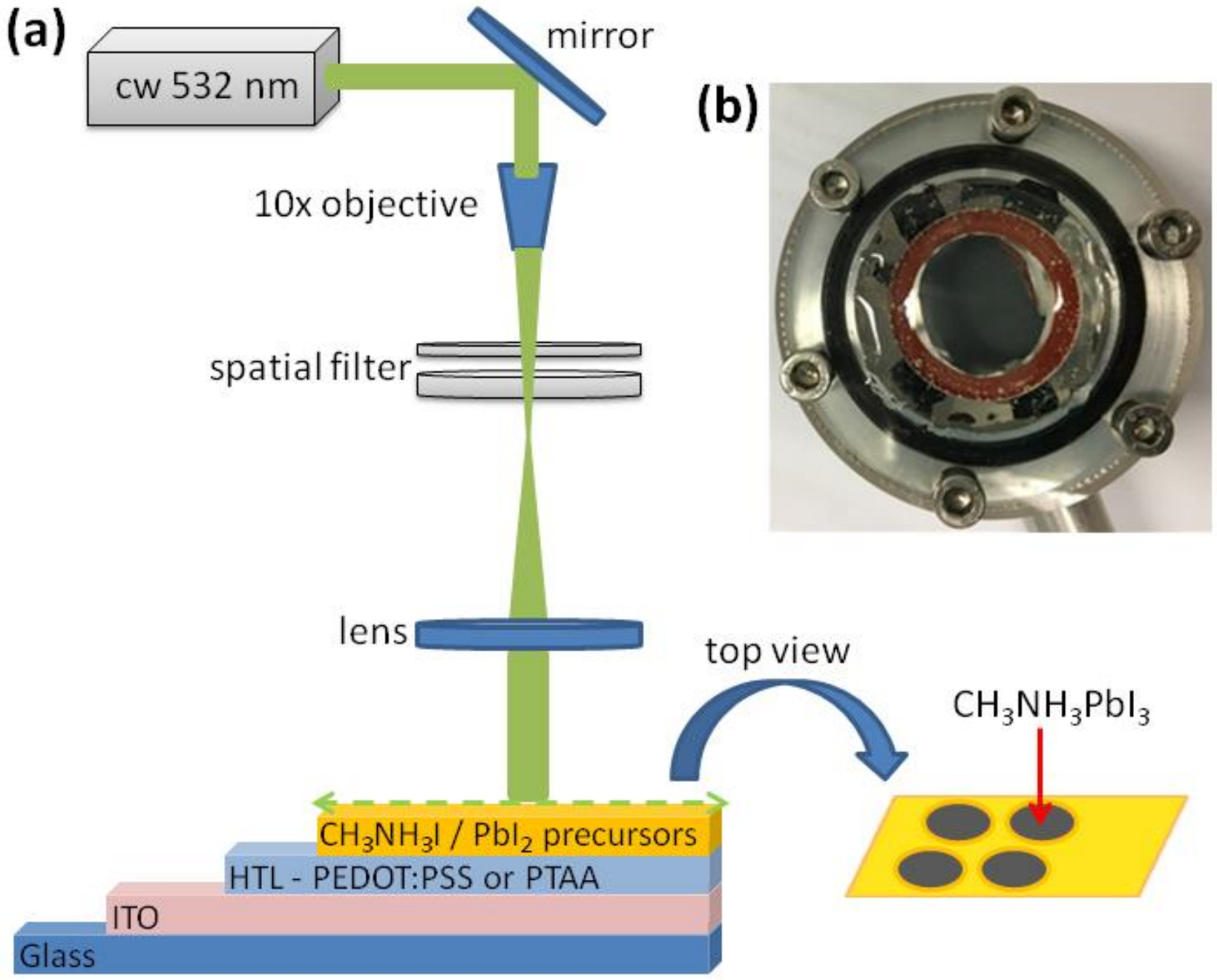

**Figure 1:** (a) Schematic representation of the laser-assisted (LA) crystallization experimental setup. (b) Homemade sample holder that allows the realization of LA crystallization and transient absorption spectroscopy (TAS) measurements under inert conditions (see text).

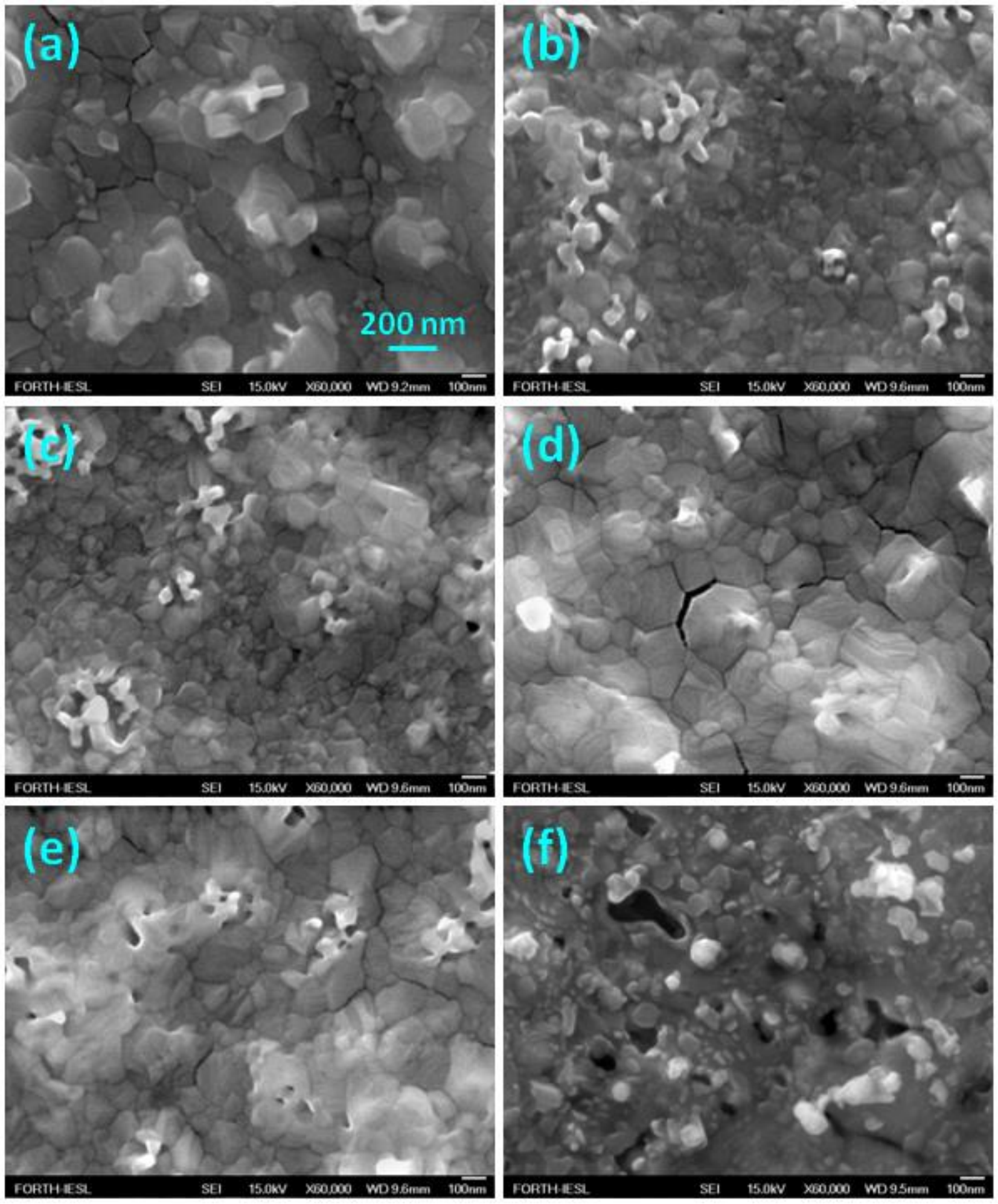

**Figure 2:** SEM images of PEDOT:PSS/$CH_3NH_3PbI_3$ films prepared following typical thermal annealing procedure **(a)**, and PEDOT:PSS/$CH_3NH_3PbI_3$ films prepared by means of laser-assisted (LA) crystallization while employing laser power densities of 6.40 W $cm^{-2}$ **(b)**, 7.65 W $cm^{-2}$ **(c)**, 8.90 W $cm^{-2}$ **(d)**, 10.15 W $cm^{-2}$ **(e)**, 11.40 W $cm^{-2}$ **(f)**. The scale bar applies to all images.

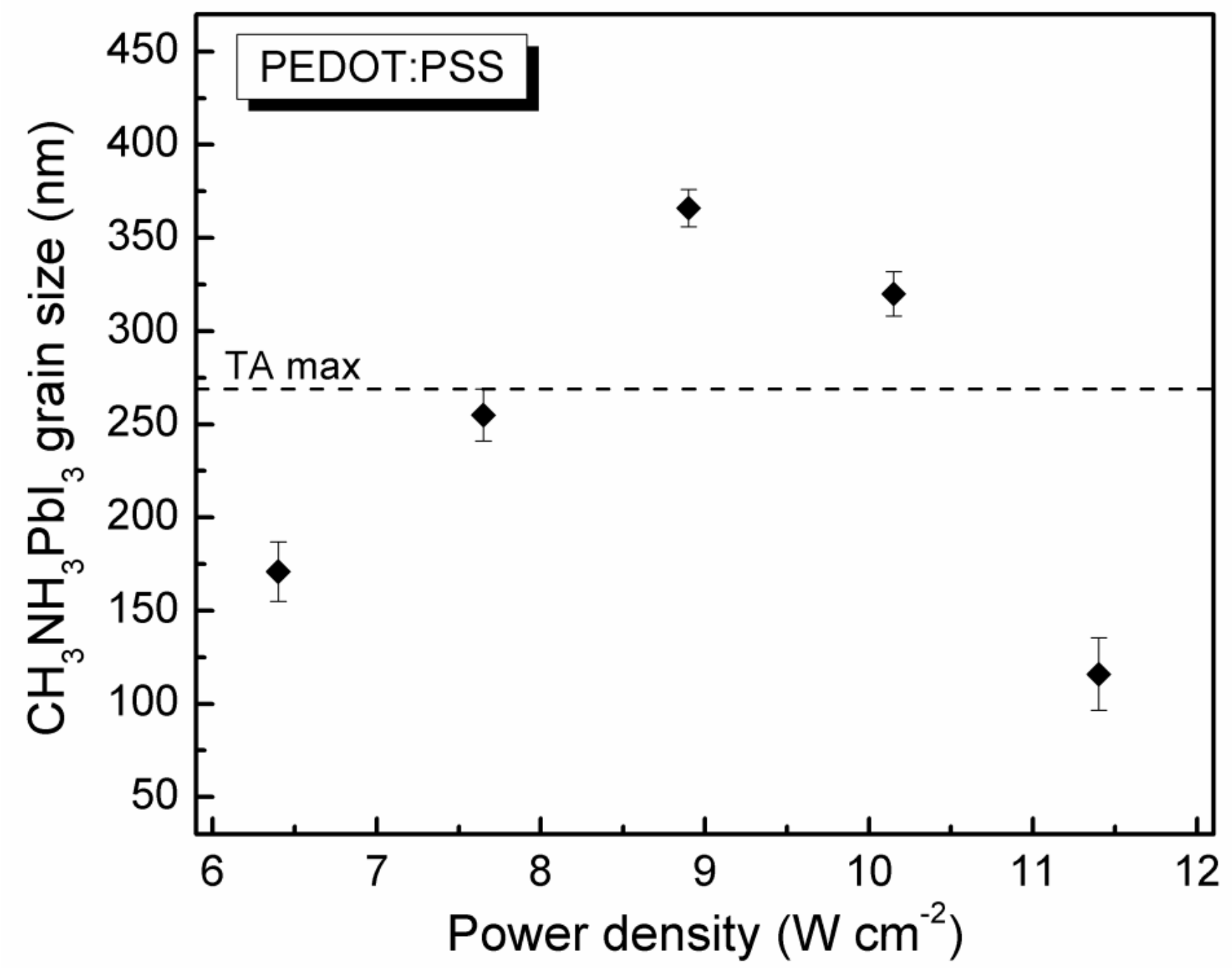

**Figure 3:** Grain size of PEDOT:PSS/$CH_3NH_3PbI_3$ films synthesized by means of laser-assisted (LA) crystallization as a function of laser power density. The grain size of the corresponding thermally annealed sample is marked by the dashed line.

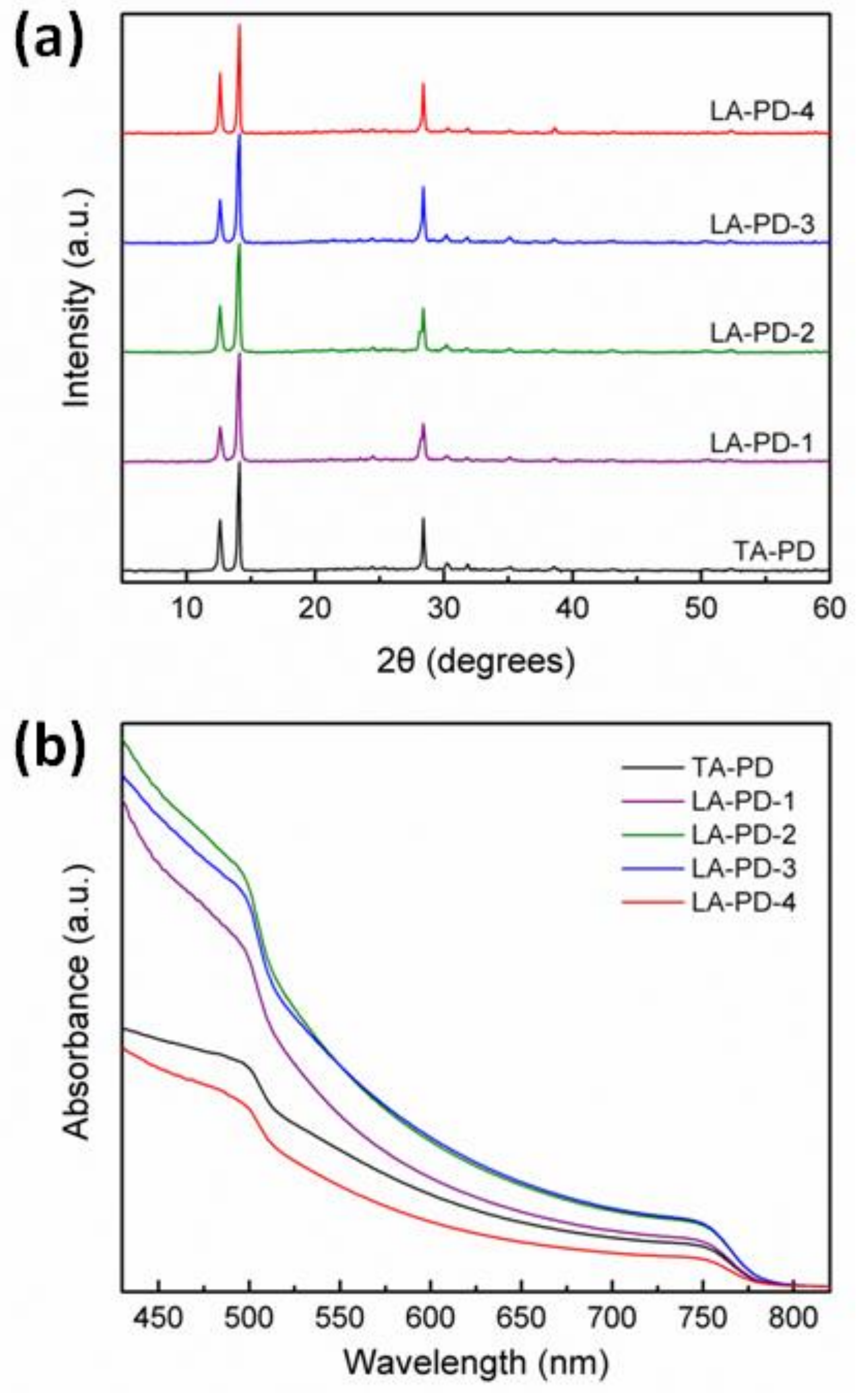

**Figure 4:** Normalized XRD patterns **(a)**, and absorbance profiles **(b)** of PEDOT:PSS/$CH_3NH_3PbI_3$ films.

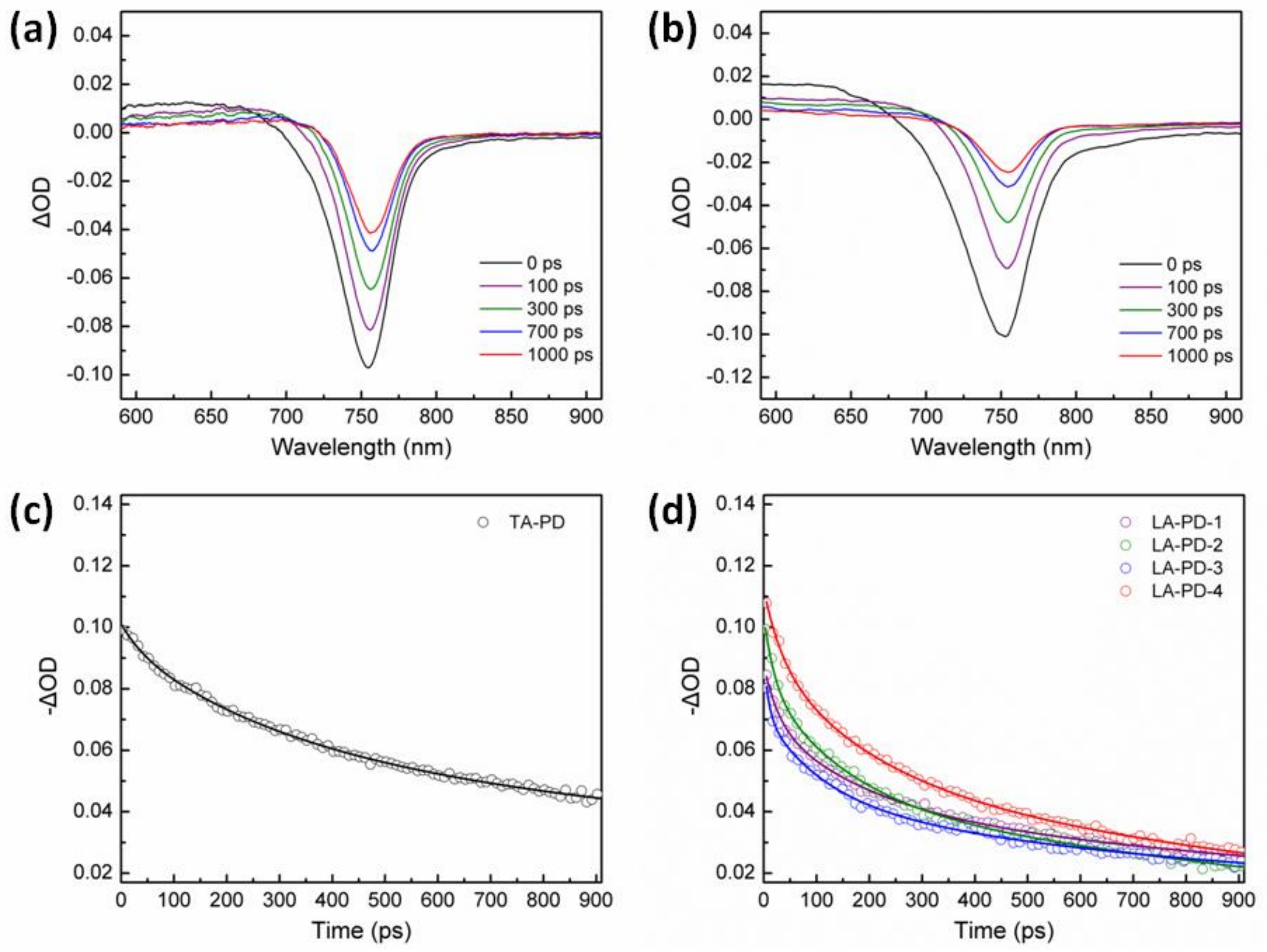

**Figure 5:** Typical ΔOD vs. wavelength plots at various time delays following photoexcitation of PEDOT:PSS/$CH_3NH_3PbI_3$ films prepared by means of thermal annealing **(a)**, and laser-assisted (LA) crystallization **(b)**. Photoexcitation was performed at 1026 nm with a pump fluence of 1 mJ $cm^{-2}$. Transient band edge bleach kinetics (empty symbols) and their corresponding decay exponential fits (lines) for the PEDOT:PSS/$CH_3NH_3PbI_3$ films prepared by means of thermal annealing **(c)**, and laser-assisted (LA) crystallization **(d)**. For fitting details see text and SI.

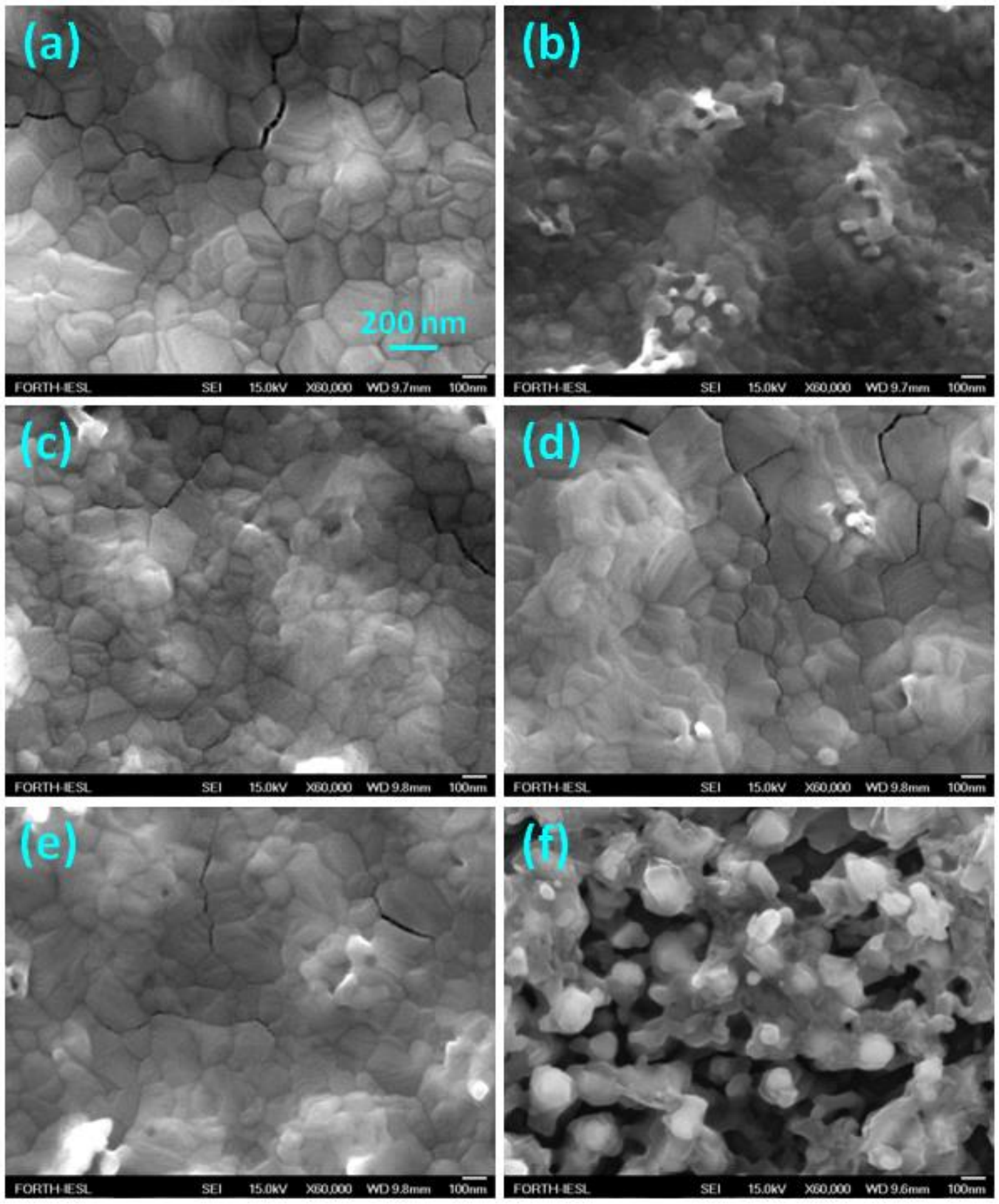

**Figure 6:** SEM images of PTAA/$CH_3NH_3PbI_3$ films prepared following typical thermal annealing procedure **(a)**, and PTAA/$CH_3NH_3PbI_3$ films prepared by means of laser-assisted (LA) crystallization while employing laser power densities of 6.40 W $cm^{-2}$ **(b)**, 7.65 W $cm^{-2}$ **(c)**, 8.90 W $cm^{-2}$ **(d)**, 10.15 W $cm^{-2}$ **(e)**, 11.40 W $cm^{-2}$ **(f)**. The scale bar applies to all images.

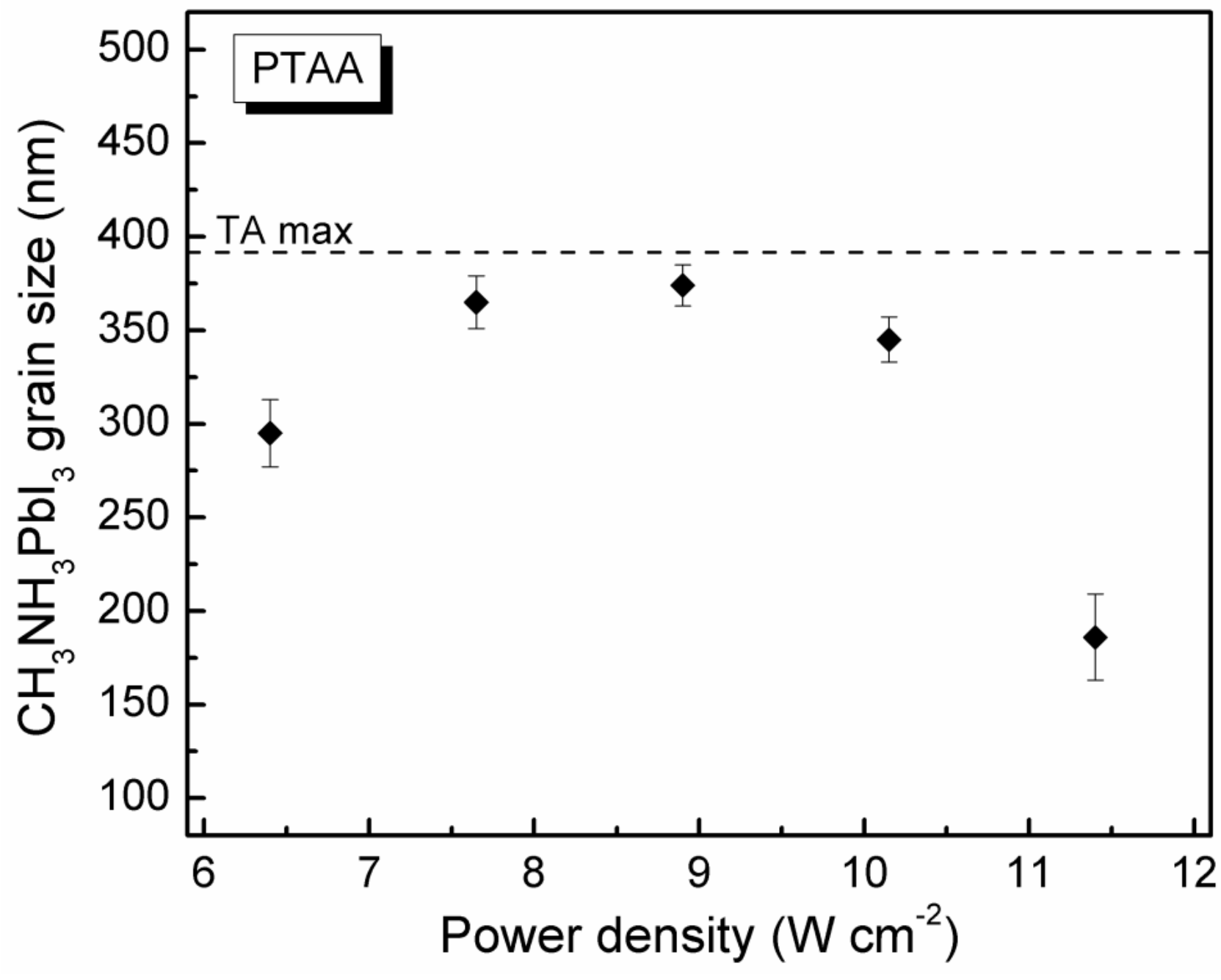

**Figure 7:** Grain size of PTAA/$CH_3NH_3PbI_3$ films synthesized by means of laser-assisted (LA) crystallization as a function of laser power density. The grain size of the corresponding thermally annealed sample is marked by the dashed line.

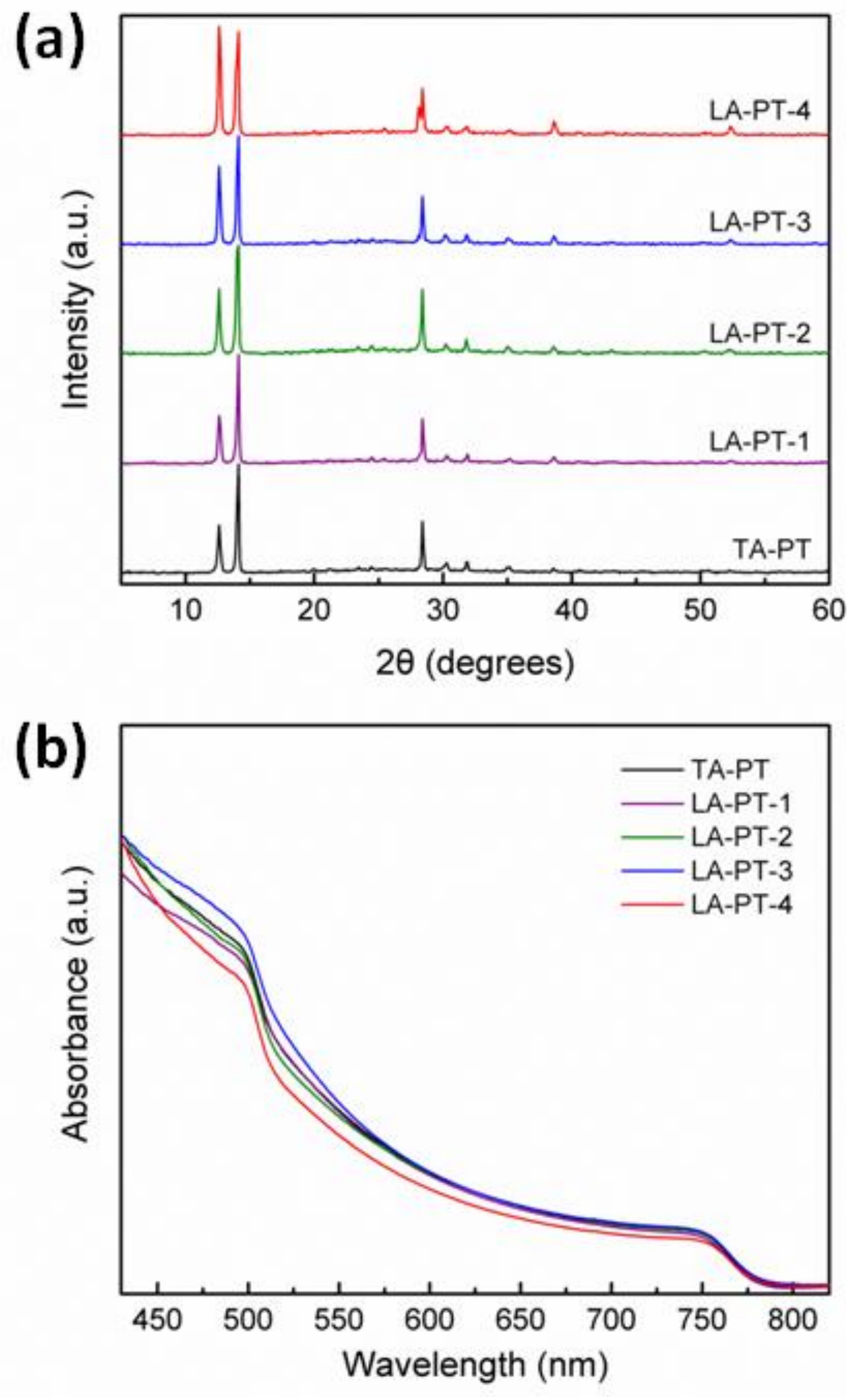

**Figure 8:** Normalized XRD patterns **(a)**, and absorbance profiles **(b)** of PTAA/$CH_3NH_3PbI_3$ films.

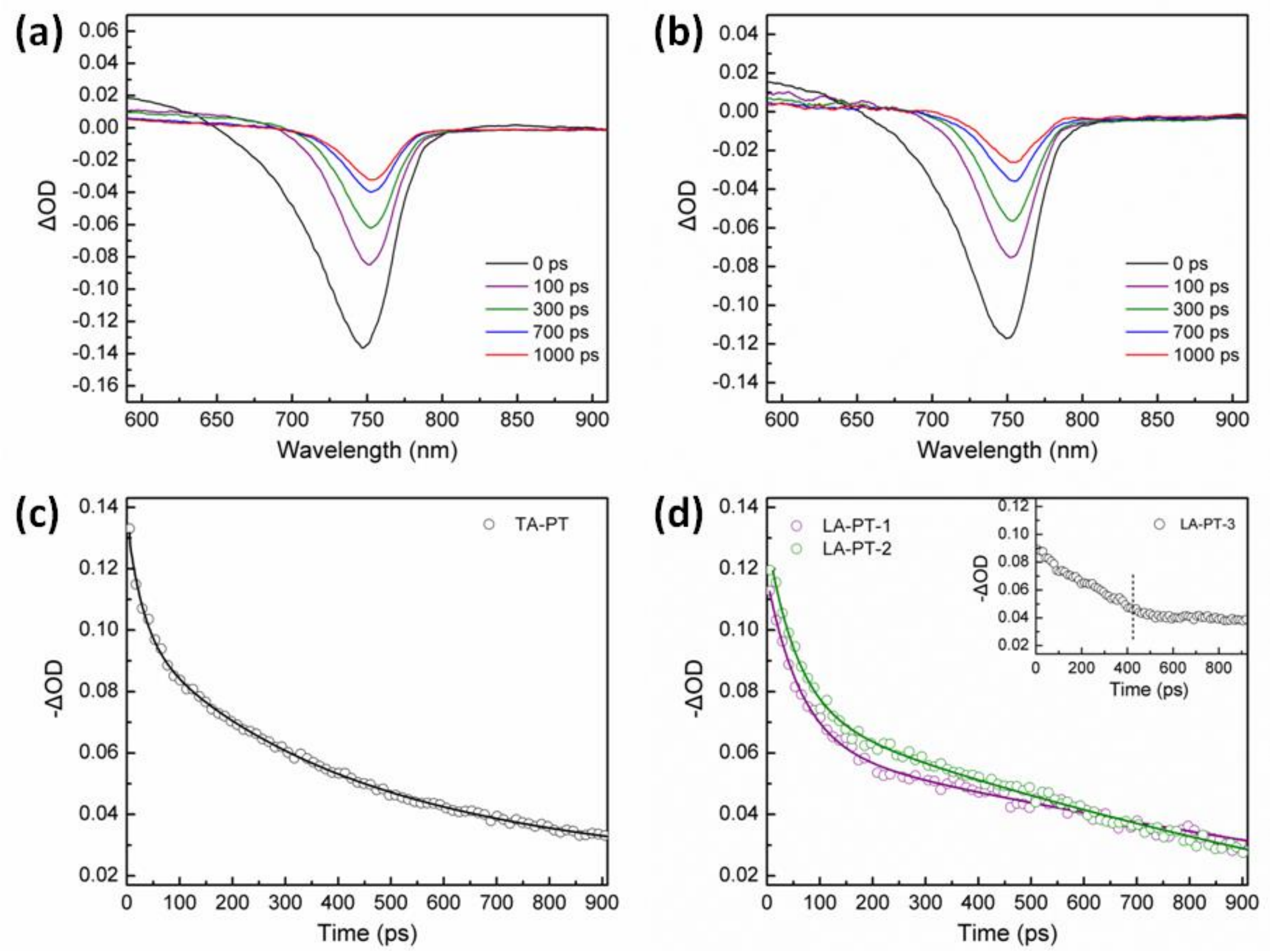

**Figure 9:** Typical ΔOD vs. wavelength plots at various time delays following photoexcitation of PTAA/$CH_3NH_3PbI_3$ films prepared by means of thermal annealing **(a)**, and laser-assisted (LA) crystallization **(b)**. Photoexcitation was performed at 1026 nm with a pump fluence of 1 mJ $cm^{-2}$. Transient band edge bleach kinetics (empty symbols) and their corresponding decay exponential fits (lines) for the PTAA/$CH_3NH_3PbI_3$ films prepared by means of thermal annealing **(c)**, and laser-assisted (LA) crystallization **(d)**. The inset shows an indicative bleach kinetics plot of a laser-assisted synthesis film that contains excessive amount of lead iodide ($PbI_2$). For fitting details see text and SI.

**Table 1:** HTL/$CH_3NH_3PbI_3$ sample configurations and corresponding synthesis parameters.

| **Sample** | **HTL polymer** | **Synthesis route** | **Conditions** |
|---|---|---|---|
| TA-PD | PEDOT:PSS | Thermal annealing | 100 ºC, 70 min |
| LA-PD-1 | PEDOT:PSS | Laser annealing | 6.40 W $cm^{-2}$, 20 s |
| LA-PD-2 | PEDOT:PSS | Laser annealing | 7.65 W $cm^{-2}$, 20 s |
| LA-PD-3 | PEDOT:PSS | Laser annealing | 8.90 W $cm^{-2}$, 20 s |
| LA-PD-4 | PEDOT:PSS | Laser annealing | 10.15 W $cm^{-2}$, 20 s |
| LA-PD-5 | PEDOT:PSS | Laser annealing | 11.40 W $cm^{-2}$, 20 s |
| TA-PT | PTAA | Thermal annealing | 100 ºC, 70 min |
| LA-PT-1 | PTAA | Laser annealing | 6.40 W $cm^{-2}$, 20 s |
| LA-PT-2 | PTAA | Laser annealing | 7.65 W $cm^{-2}$, 20 s |
| LA-PT-3 | PTAA | Laser annealing | 8.90 W $cm^{-2}$, 20 s |
| LA-PT-4 | PTAA | Laser annealing | 10.15 W $cm^{-2}$, 20 s |
| LA-PT-5 | PTAA | Laser annealing | 11.40 W $cm^{-2}$, 20 s |

**Table 2:** Time components for the HTL/$CH_3NH_3PbI_3$ configurations following exponential fitting (see text).

| **Sample** | **$\lambda_{max}$ (nm)** | **$\tau_1$ (ps)** | **$\tau_2$ (ps)** | **$\tau_3$ (ps)** |
|---|---|---|---|---|
| TA-PD | 754 | 62 | 440 | $5.85 \times 10^5$ |
| LA-PD-1 | 752 | 28 | 218 | $6.30 \times 10^6$ |
| LA-PD-2 | 752 | 20 | 171 | $8.74 \times 10^6$ |
| LA-PD-3 | 752 | 10 | 111 | $1.32 \times 10^7$ |
| LA-PD-4 | 752 | 34 | 224 | $4.80 \times 10^6$ |
| TA-PT | 750 | 31 | 375 | $1.14 \times 10^5$ |
| LA-PT-1 | 751 | 64 | 2256 | $3.20 \times 10^5$ |
| LA-PT-2 | 751 | 57 | 1813 | $5.10 \times 10^5$ |
| LA-PT-3 | 751 | - | - | - |
| LA-PT-4 | 751 | - | - | - |

**Table 3:** Charge carrier recombination rate constants for the HTL/$CH_3NH_3PbI_3$ configurations following polynomial rate equation fitting (see text).

| **Sample** | **$\lambda_{max}$ (nm)** | **$k_3$ ($cm^6$ $s^{-1}$) ± 0.5** | **$k_2$ ($cm^3$ $s^{-1}$) ± 0.3** | **$k_1$ ($\mu s^{-1}$) ± 0.2** |
|---|---|---|---|---|
| TA-PD | 754 | $9.6 \times 10^{-14}$ | $2.6 \times 10^{-9}$ | $2.8 \times 10^{-7}$ |
| LA-PD-1 | 752 | $2.4 \times 10^{-13}$ | $7.9 \times 10^{-10}$ | $5.5 \times 10^{-7}$ |
| LA-PD-2 | 752 | $3.5 \times 10^{-13}$ | $8.6 \times 10^{-10}$ | $7.7 \times 10^{-7}$ |
| LA-PD-3 | 752 | $2.5 \times 10^{-13}$ | $6.2 \times 10^{-10}$ | $5.6 \times 10^{-7}$ |
| LA-PD-4 | 752 | $3.2 \times 10^{-13}$ | $8.0 \times 10^{-10}$ | $7.4 \times 10^{-7}$ |
| TA-PT | 752 | $3.7 \times 10^{-13}$ | $8.9 \times 10^{-10}$ | $8.0 \times 10^{-7}$ |
| LA-PT-1 | 751 | $7.0 \times 10^{-13}$ | $1.6 \times 10^{-9}$ | $1.3 \times 10^{-6}$ |
| LA-PT-2 | 751 | $7.1 \times 10^{-13}$ | $1.3 \times 10^{-9}$ | $1.1 \times 10^{-6}$ |
| LA-PT-3 | 751 | - | - | - |
| LA-PT-4 | 751 | - | - | - |

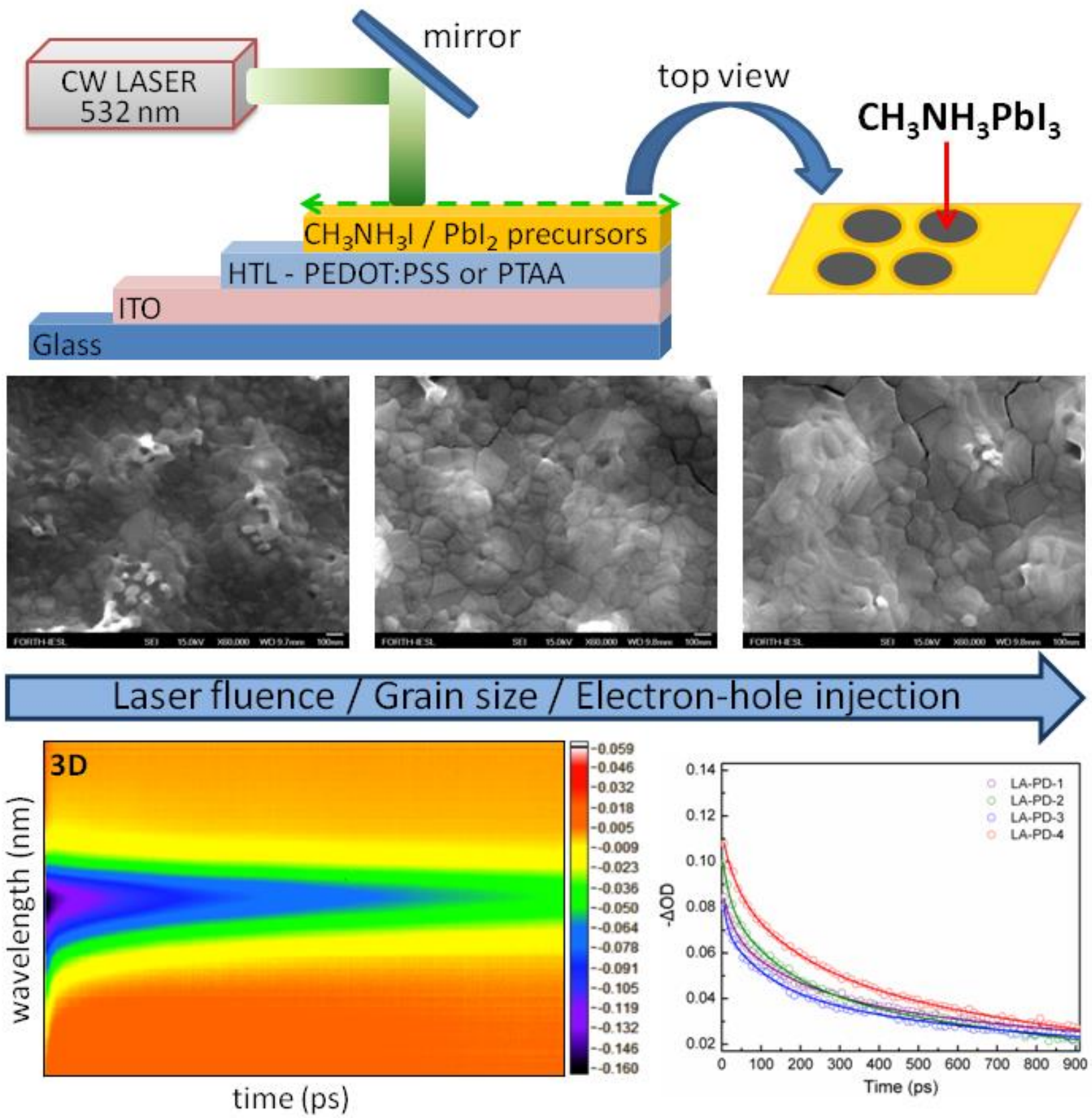

**Table of Content (ToC) Figure:** Charge carrier dynamics of $CH_3NH_3PbI_3$ perovskite films synthesized by means of laser-assisted crystallization on two types of hole transport layer (HTL) polymers.